\documentclass[aps,prd,preprintnumbers,superscriptaddress,nofootinbib,notitlepage,floatfix,10pt]{revtex4-2}
\usepackage[pdftex]{graphicx}
\usepackage{bm,latexsym,amsmath,amssymb,amsfonts,mathrsfs}
\usepackage{booktabs} % Allows the use of \toprule, \midrule and \bottomrule in tables for horizontal lines
\usepackage{color}
\allowdisplaybreaks[1]
\usepackage[pdftex,colorlinks=true,linkcolor=blue,citecolor=cyan,backref=page]{hyperref}
\newcommand*{\D}{\mathrm{d}}
\newcommand*{\mpl}{M_{\mathrm{Pl}}}
\begin{document}
%-----------------------------------------------------------------%
\title{Weak-field regime of scalar-tensor theories with an instantaneous mode}
%-----------------------------------------------------------------%
%
\author{Tsutomu~Kobayashi}
\email[Email: ]{tsutomu@rikkyo.ac.jp}
\affiliation{Department of Physics, Rikkyo University, Toshima, Tokyo 171-8501, Japan}
\author{Takashi~Hiramatsu}
\email[Email: ]{hiramatz@rikkyo.ac.jp}
\affiliation{Department of Physics, Rikkyo University, Toshima, Tokyo 171-8501, Japan}
%
%-----------------------------------------------------------%
\begin{abstract}
Higher-order scalar-tensor theories having an instantaneous mode do not develop the Ostrogradsky instability even if a seemingly dangerous mode is present.
Such theories satisfy only partially the degeneracy conditions that are usually imposed to remove
the dangerous mode completely, and are dubbed as U-DHOST theories.
We study weak gravitational fields sourced by nonrelativistic matter distributions in U-DHOST theories.
In contrast to the case of totally degenerate theories where nonlinear derivative interactions
are crucial for exhibiting the Vainshtein mechanism and its partial breaking,
we show that in generic U-DHOST theories the linear analysis is sufficient for weak gravitational fields.
We identify the subset of U-DHOST theories in which solar system tests are evaded and
gravitational waves propagate at the speed of light.
Such theories can however be tested with cosmological observations.
We also point out that there is a particular case of U-DHOST theories where nonlinear derivative interactions play an important role as in totally degenerate theories. In that case, it is found, however, that the Vainshtein mechanism does not operate and gravitational forces are modified.
\end{abstract}
%-----------------------------------------------------------------%
\preprint{RUP-23-21}
\maketitle
%-----------------------------------------------------------------%

\section{Introduction}\label{sec:intro}

The great interest in the mysterious mechanism for driving the accelerated expansion
of the present Universe has been stimulating the study of modified gravity over the years.
Adding a scalar degree of freedom on top of the tensorial (gravitational-wave) degrees
of freedom is the simplest way of modifying general relativity.
Since the rediscovery~\cite{Deffayet:2011gz,Kobayashi:2011nu} of the Horndeski theory~\cite{Horndeski:2022xng},
a considerable effort has been devoted to attempts
to generalize scalar-tensor theories of gravity without
evoking the dangerous Ostrogradsky ghosts (see, e.g., Refs.~\cite{Langlois:2018dxi,Kobayashi:2019hrl} for a review).
The Horndeski theory is the most general scalar-tensor theory with second-order
field equations, and hence trivially evades the Ostrogradsky ghosts.
A naive expectation is that extending the Horndeski theory to higher-derivative field equations
would inevitably lead to Ostrogradsky ghosts. However, one still has healthy
scalar-tensor theories with only one scalar and two tensorial degrees of freedom
if the system is degenerate. The first example that exploits this loophole is
obtained by applying a derivative-dependent disformal transformation to the
Horndeski theory~\cite{Zumalacarregui:2013pma}. Rewriting the Horndeski action
by the use of the Arnowitt-Deser-Misner (ADM) formulation and then liberating the coefficient
of each term from the Horndeski tuning, one can arrive at a degenerate theory
relatively easily~\cite{Gleyzes:2014dya}.
Systematic constructions and classifications of degenerate higher-order
scalar-tensor (DHOST) theories
are presented in Refs.~\cite{Langlois:2015cwa,Crisostomi:2016czh,BenAchour:2016fzp}.
More recently, novel higher-order scalar-tensor theories were derived by applying
to the Horndeski theory a generalized disformal transformation containing yet higher
derivatives~\cite{Takahashi:2021ttd,Takahashi:2022mew,Takahashi:2023vva}.
The consistency of matter couplings in the theories generated via generalized disformal transformation is nontrivial and has been investigated in Refs.~\cite{Deffayet:2020ypa,Naruko:2022vuh,Ikeda:2023ntu}.

In DHOST theories, a set of conditions are imposed among the otherwise arbitrary functions
in the Lagrangian to make the system degenerate and thus remove would-be dangerous Ostrogradsky modes.
The point here is that the degeneracy conditions are introduced so that
the system is degenerate irrespective of the choice of time slicing.
Recently, a further generalization of higher-order scalar-tensor theories has been proposed
in which the degeneracy conditions are satisfied only when the scalar degree of freedom is
homogeneous on constant time hypersurfaces, i.e. in the unitary gauge~\cite{DeFelice:2018ewo}.
Such relaxed versions of degenerate theories are dubbed as U-DHOST theories.
In U-DHOST theories, an apparent additional
degree of freedom appears away from the unitary gauge. However, this degree of freedom turns out to be
an instantaneous mode (a nonpropagating ``shadowy mode'' in the terminology of Refs.~\cite{DeFelice:2018ewo,DeFelice:2021hps}) satisfying an elliptic equation on spatial hypersurfaces rather than a hyperbolic equation. Its configuration is therefore completely determined by boundary conditions and hence a shadowy mode is harmless.
Interestingly, U-DHOST theories can resolve the strong coupling problem of
stealth solutions in DHOST theories~\cite{DeFelice:2022xvq}.
Furthermore, U-DHOST theories have been used to reformulate the action for DHOST theories
in Ref.~\cite{Langlois:2020xbc}.

The purpose of the present paper is to study a phenomenological aspect of such higher-order scalar-tensor theories satisfying the degeneracy conditions only partially (i.e. only in the unitary gauge) and
having an instantaneous mode. More specifically, we study weak gravitational fields
produced by a nonrelativistic source in U-DHOST theories and highlight the difference from the case of
totally degenerate theories. In Horndeski and DHOST theories, nonlinear derivative interaction terms play a crucial role in determining a weak-field configuration.
In the (quadratic) Horndeski theory, the scalar degree of freedom is screened and the standard behavior of gravity is recovered in the vicinity of a source due to this nonlinear effect~\cite{Kimura:2011dc,Narikawa:2013pjr,Koyama:2013paa}, which is called the Vainshtein mechanism.
In generic (quadratic) DHOST theories, while screening is complete in the exterior region of a source, a partial breaking of the Vainshtein mechanism occurs and gravitational forces are modified in the matter interior~\cite{Kobayashi:2014ida,Dima:2017pwp,Langlois:2017dyl}.
There is a particular subset of DHOST theories in which the breaking of screening occurs in a different way~\cite{Hirano:2019scf,Crisostomi:2019yfo}.
In this paper, we extend these previous works to U-DHOST theories.

The paper is organized as follows. In the next section, we briefly review U-DHOST theories.
Then, in Sec.~\ref{sec:Vainshtein}, we derive the effective action governing
weak gravitational fields in U-DHOST theories.
We introduce so-called EFT parameters, emphasizing how the degeneracy conditions
in DHOST theories are relaxed in U-DHOST theories in terms of the EFT parameters.
In Sec.~\ref{sec:LR}, we consider generic U-DHOST theories and show that a linear analysis is sufficient for weak gravitational fields as opposed to the case of fully degenerate theories in which nonlinear derivative interactions play an important role.
On the basis of the linear analysis, we identify a subset of U-DHOST theories that evades experimental tests.
However, the analysis performed in Sec.~\ref{sec:LR} is not valid if the functions in the Lagrangian satisfy a certain relation. This special case is investigated in some detail in Sec.~\ref{sec:VR}.
We draw our conclusions in Sec.~\ref{sec:concl}.

\section{Quadratic U-DHOST theories}\label{sec:UDHOSTtheories}

Using the notations $\phi_\mu=\nabla_\mu\phi$ and $\phi_{\mu\nu}=\nabla_\mu\nabla_\nu\phi$
for derivatives of the scalar field $\phi$,
the action for quadratic U-DHOST theories~\cite{DeFelice:2018ewo} is written as
\begin{align}
        S&=\int\D^4x\sqrt{-g}\left[P+Q\Box\phi+fR+\sum_{I=1}^5A_IL_I+\mathcal{L}_\textrm{m}\right],\label{action:udhost}
\end{align}
with 
\begin{align}
        L_1&:=\phi_{\mu\nu}\phi^{\mu\nu},\quad 
        L_2:=(\Box\phi)^2,\quad L_3:=\Box\phi\phi^\mu\phi_{\mu\nu}\phi^\nu,
        \quad 
        L_4:=\phi^\mu\phi_{\mu\nu}\phi^{\nu\lambda}\phi_\lambda,
        \quad L_5:=\left(\phi^\mu\phi_{\mu\nu}\phi^\nu\right)^2,\label{lag:sup1}
\end{align}
where $P$, $Q$, and $f$ are arbitrary functions of $\phi$ and $X:=-g^{\mu\nu}\phi_\mu\phi_\nu/2$,
and $R$ is the Ricci scalar.
The five coefficients $A_1,\dots,A_5$ are also functions of $\phi$ and $X$,
and they are expressed in terms of $f$ and four arbitrary functions $a_1,\dots,a_4$ of $\phi$ and $X$ as
\begin{align}
        &A_1=a_1-\frac{f}{2X},
        \quad 
        A_2=a_2+\frac{f}{2X},
        \quad 
        A_3=\frac{f}{2X^2}-\frac{f_X}{X}+2a_1a_3+
        2\left(3a_3+\frac{1}{2X}\right)a_2,
        \notag \\ &
        A_4=a_4+\frac{f_X}{2X}-\frac{f}{2X^2}+\frac{a_1}{X},
        \quad 
        A_5=\frac{a_4}{2X}-\frac{f_{X}}{4X^2}+a_1 
        \left(\frac{1}{4X^2}+3a_3^2+\frac{a_3}{X}\right)
        +a_2\left(3a_3+\frac{1}{2X}\right)^2.\label{lag:sup2}
\end{align}
Therefore, five of the six functions ($f,A_1,\dots, A_5$) are independent.
We include the Lagrangian for minimally coupled matter $\mathcal{L}_\textrm{m}$.
In this paper, this is taken to be nonrelativistic matter.

U-DHOST theories are degenerate only in the unitary gauge
and do not satisfy all the degeneracy conditions imposed on totally degenerate theories.
In the following cases, the above action reduces to that of DHOST theories~\cite{Langlois:2015cwa}.
The class Ia degeneracy conditions are satisfied if 
\begin{align}
        a_1&=-a_2\neq 0,
        \\ 
        a_4&=2a_3\left(\frac{1}{X}-a_3\right)f+\left(\frac{1}{2X}-4a_3\right)f_X,\label{deg-Ia-2}
\end{align}
where $f_X:=\partial f/\partial X$ (and in what follows we will use the same notation for the other functions).
In this class, only $a_2$, $a_3$, and $f$ are arbitrary.
The class Ia theories are generated from the Horndeski theory by performing a disformal transformation~\cite{BenAchour:2016cay}.
The class IIa degeneracy conditions are satisfied if
\begin{align}
        a_3&=\frac{f-2Xf_X}{2Xf},
        \\ 
        a_4&=\frac{f^2-3Xff_X+4X^2f_X^2}{2X^2f}.
\end{align}
In this case, $a_1\,(\neq 0)$, $a_2\,(\neq -a_1)$, and $f\,(\neq 0)$ are arbitrary.
The class IIa theories are disformally disconnected from the Horndeski theory~\cite{BenAchour:2016cay}, and cosmological solutions in this class of theories
exhibit instabilities either in the scalar or tensor sector~\cite{deRham:2016wji,Langlois:2017mxy}.
In this paper, we are interested in the phenomenology of U-DHOST theories, and hence
do \textit{not}
impose the above additional conditions that force the system to degenerate in any gauge.

We will expand the action~\eqref{action:udhost} in perturbations around a cosmological background
with $\phi=\phi_0(t)$.
For such an analysis, it is convenient to introduce the effective Planck mass $M$ defined by
\begin{align}
    M^2:=4Xa_1
\end{align}
and the dimensionless parameters
$\alpha_T$, $\alpha_H$, $\beta_1$, $\beta_3$, $\alpha_L$, $\alpha_{V1}$, and $\alpha_{V2}$
(the so-called EFT of dark energy parameters)
defined by\footnote{The parameters $\alpha_{V1}$ and $\alpha_{V2}$ here are different from those introduced in Ref.~\cite{Cusin:2017mzw}.}
\begin{align}
        &M^2\alpha_T=2(f-2Xa_1),
        \quad 
        M^2\alpha_H=2(f-2Xf_X-2Xa_1),
        \notag \\ 
        &M^2\beta_1:%=2X\left[f_X-A_2+XA_3\right]
        =4X^2\left(a_1+3a_2\right)a_3,
        \quad 
        M^2\beta_2 %=4X\left[A_1+A_2-2X(A_3+A_4)+4X^2A_5\right]
        :=48X^3(a_1+3a_2)a_3^2,
        \quad  
        M^2\beta_3:%=-8X\left[f_X+A_1-XA_4\right]
        =-4X\left(f_X-2Xa_4\right),
        \notag \\ 
        &M^2\alpha_L:%=-6X(A_1+A_2)
        =-6X(a_1+a_2),
        \quad
        M^2\alpha_{V1}:=8X(a_1+Xa_{1X}),\quad M^2\alpha_{V2}:=-8X(a_2+Xa_{2X}),
        \label{defs:eftparas}
\end{align}
where the right-hand side quantities are evaluated at $\phi=\phi_0(t)$
and hence these parameters are functions of time.
We will assume that the dimensionless EFT parameters are of order $\mathcal{O}(1)$
(or smaller) when they are nonvanishing.
Note that $\beta_2$ can be expressed in terms of the other parameters as $\beta_2=-6\beta_1^2/(1+\alpha_L)$.
See Appendix~\ref{app:eftlag} for some details on how the EFT parameters are defined.

The degeneracy conditions in class Ia DHOST theories impose
\begin{align}
        \alpha_L=0,\quad \alpha_{V1}=\alpha_{V2}, \quad 
        \beta_3=-2\beta_1[2(1+\alpha_H)+(1+\alpha_T)\beta_1],\label{deg-Ia-para}
\end{align}
while those in class IIa DHOST theories read
\begin{align}
        \alpha_L\neq 0, \quad \beta_1=-\frac{(1+\alpha_L)(1+\alpha_H)}{1+\alpha_T},
        \quad 
        \beta_3=\frac{2(1+\alpha_H)^2}{1+\alpha_T}.\label{deg-IIa-para}
\end{align}
In U-DHOST theories, we need to require none of these conditions, and
hence all the EFT parameters are in principle arbitrary
(except for $\beta_2$, which is subject to the aforementioned relation).

One might be interested in theories in which
the propagation speed of gravitational waves, $c_\textrm{GW}$, is equal to that of light.
The gravitational-wave sector has nothing to do with the degeneracy conditions, and
it is easy to see that $c_\textrm{GW}^2=f/(2Xa_1)$.
In terms of the EFT parameters, the conditions under which $c_\textrm{GW}=1$ read
\begin{align}
        \alpha_T=0,\quad \alpha_H=-\alpha_{V1}.
\end{align}
Note, on the other hand, that there is a subtlety when applying
the LIGO bound on the speed of gravitational waves to modified gravity as an
alternative to dark energy~\cite{deRham:2018red}.

\section{Effective action for weak gravitational fields}\label{sec:Vainshtein}

%---------------------------------------------------------------------%
%\renewcommand{\arraystretch}{2.2}% Parameter for the line width; Change the value appropriately.
\begin{table}[tb]
\begin{tabular}{cc}
\toprule
 Quantity & Order  \\ 
 \midrule
 $\dot\phi_0$ & $\mpl H_0$
 \\
 \midrule
 $M$ & $\mpl$
 \\ 
 $\alpha_T,\alpha_H,\alpha_L,\alpha_{V1},\alpha_{V2},\beta_1,\beta_3$ 
 & 1
 \\ 
 \midrule
 $c_1,c_2$ & $M^{-1}$ 
 \\
 $c_3$ & $M^{-2}$
 \\ 
 $c_4, c_5, c_6$ & $1$
 \\ 
 $c_7,c_8$ & $M^{-1}H_0^{-1}$
 \\ 
 $c_9, c_{10}$ & $M^{-2}H_0^{-2}$
 \\
 $b_1,\widetilde{b}_1$ & $M^{-3}H_0^{-2}$
 \\ 
 $b_2,\widetilde{b}_2,b_3,\widetilde{b}_3,b_4,b_5$ & $M^{-2}H_0^{-2}$
 \\ 
 $b_6,\widetilde{b}_6$ & $M^{-3}H_0^{-3}$
 \\ 
 $d_1,\widetilde{d}_1,d_2$ & $M^{-4}H_0^{-4}$
 \\
\bottomrule
\end{tabular}
\caption{Orders of magnitude for various quantities.}
\label{table:order-list}
\end{table}
%---------------------------------------------------------------------%

We now derive the action governing weak gravitational fields around a cosmological background 
in U-DHOST theories. The metric in the Newtonian gauge is written as
\begin{align}
        \D s^2=-[1+2\Phi(t,\Vec{x})]\D t^2+a^2(t)[1-2\Psi(t,\Vec{x})]\D\Vec{x}^2,
\end{align}
and the scalar field is given by 
\begin{align}
        \phi=\phi_0(t)+\pi(t,\Vec{x}).
\end{align}
We expand the action~\eqref{action:udhost} in perturbations, keeping nonlinear terms with possibly large spatial gradients $\nabla$ on scales much smaller than the Hubble horizon scale.
To do so, it is convenient to assign a small bookkeeping parameter $\varepsilon$ to the perturbations $\Phi$, $\Psi$, and $\pi$.
Keeping the terms containing at least two spatial derivatives,
the perturbative expansion of the Lagrangian for quadratic U-DHOST theories yields
\begin{align}
    \mathcal{L}&\sim \nabla^2\varepsilon^2,\quad
    \nabla^4\varepsilon^2,
    \notag \\ & \quad 
    \nabla^2\varepsilon^3,\quad \underline{\nabla^4\varepsilon^3},
    \notag \\ & \quad 
    \nabla^2\varepsilon^4,\quad \nabla^4\varepsilon^4,\quad \underline{\nabla^6\varepsilon^4},
    \notag \\ & \quad 
    \nabla^2\varepsilon^5,\quad \nabla^4\varepsilon^5,\quad \nabla^6\varepsilon^5,
    \notag \\ & \quad 
    \nabla^2\varepsilon^6,\quad \nabla^4\varepsilon^6,\quad \nabla^6\varepsilon^6,
    \quad \nabla^8\varepsilon^6,
    \notag \\ & \quad 
    \nabla^2\varepsilon^7,\quad \nabla^4\varepsilon^7,\quad \nabla^6\varepsilon^7,
    \quad \nabla^8\varepsilon^7,
    \notag \\ & \quad 
    \nabla^2\varepsilon^8,\quad \nabla^4\varepsilon^8,\quad \nabla^6\varepsilon^8,
    \quad \nabla^8\varepsilon^8,\quad \nabla^{10}\varepsilon^8,
    \notag \\ & \quad 
    \nabla^2\varepsilon^9, \quad \dots,
\end{align}
where the first line gives linear terms in the field equations and the other lines show nonlinearities.
(From the structure of the Lagrangian of quadratic U-DHOST theories we see that
there appears no term of the form $\nabla^{2(n-1)}\varepsilon^n$ with $n\ge 5$.)
Among those nonlinear terms, the underlined two terms contain the largest numbers of spatial derivatives per field and we assume that they can be as large as the terms in the first line.
It can then be seen that all the other nonlinear terms are smaller.
The situation here is essentially the same as that in Horndeski and DHOST theories.
The only new ingredient in U-DHOST theories is the term in the first line that scales as $\sim 
\nabla^4\varepsilon^2$.

%Keeping all such nonlinear higher-derivative terms, we obtain

We thus keep the terms that scale as $\nabla^2\varepsilon^2$,
$\nabla^4\varepsilon^2$, $\nabla^4\varepsilon^3$, and $\nabla^6\varepsilon^4$
in the perturbative expansion, and obtain
\begin{align}
        \mathcal{L}_{\textrm{eff}}&=\frac{M^2a}{2}\biggl[
                \left(
                        c_1\Phi+c_2\Psi+c_3\pi
                \right)\nabla^2\pi + c_4\Psi\nabla^2\Phi +c_5 \Psi\nabla^2\Psi 
                +c_6\Phi\nabla^2\Phi 
                %\notag \\ & \quad 
                +\left(c_7\dot\Psi+c_8\dot\Phi+c_9\ddot\pi\right)\nabla^2\pi 
                +\frac{c_{10}}{a^2}(\nabla^2\pi)^2
                \notag \\ & \quad 
                +\frac{b_1}{a^2}\mathcal{L}_3^{\textrm{Gal}}
                +\frac{\widetilde{b}_1}{a^2}\pi(\nabla^2\pi)^2 
                +\frac{1}{a^2}\left(b_2 \Phi+b_3\Psi \right)\mathcal{E}_3^{\textrm{Gal}}
                +\frac{1}{a^2}\left(\widetilde{b}_2 \Phi+\widetilde{b}_3\Psi \right)
                \mathcal{E}_3^{+}
                \notag \\ & \quad 
                +\frac{1}{a^2}\left(b_4\nabla_i\Psi+b_5\nabla_i\Phi +b_6\nabla_i\dot\pi \right) 
                \nabla_j\pi\nabla_i\nabla_j\pi 
                +\frac{\widetilde{b}_6}{a^2}\dot\pi (\nabla^2\pi)^2
                \notag \\ & \quad 
                +\frac{1}{a^4}\left(
                        d_1\mathcal{L}_4^{\textrm{Gal}}
                        +\widetilde{d}_1\widetilde{\mathcal{L}}_4
                        +d_2\nabla_i\pi\nabla_j\pi \nabla_i\nabla_k\pi \nabla_j\nabla_k\pi 
                        \right)
        \biggr]-a^3\Phi \rho,\label{effLag}
\end{align}
where
\begin{align}
        &\mathcal{E}_3^{\textrm{Gal}}:=(\nabla^2\pi)^2-\nabla_i\nabla_j\pi\nabla_i\nabla_j\pi,
        \quad  
        \mathcal{E}_3^{+}:=(\nabla^2\pi)^2+\nabla_i\nabla_j\pi\nabla_i\nabla_j\pi,
        \notag \\ &
        \mathcal{L}_3^{\textrm{Gal}}:=-\frac{1}{2}(\nabla\pi)^2\nabla^2\pi,
        \quad 
        \mathcal{L}_4^{\textrm{Gal}}:=-\frac{1}{2}(\nabla\pi)^2\mathcal{E}_3^{\textrm{Gal}},
        \quad 
        \widetilde{\mathcal{L}}_4:=-\frac{1}{2}(\nabla\pi)^2\mathcal{E}_3^{+},
\end{align}
and $\rho=\rho(t,\Vec{x})$ is the nonrelativistic matter overdensity.\footnote{It may therefore be appropriate
to write this quantity as $\delta\rho(t,\Vec{x})$.}
Here, a dot denotes differentiation with respect to $t$.
The explicit expressions for the coefficients $c_1,c_2,\dots$ in terms of the EFT parameters are given in Appendix~\ref{app:cff}.
The coefficients $c_{10}$, $\tilde{b}_i$, and $\tilde{d}_1$ vanish in the case of class Ia totally degenerate theories.
%It is customary to introduce another
%parameter $\beta_2:=48X^3(a_1+3a_2)a_3^2=-6\beta_1^2/(1+\alpha_L)$, which does not appear
%explicitly in the Lagrangian~\eqref{effLag}.
%(See also Appendix~\ref{app:eftlag}.)
%Note that $M^2\neq 0$ is assumed throughout the paper.
%The other coefficients $c_1$, $c_2$, $c_3$, and $b_1$ cannot be written in terms of
%the above parameters and have much more complicated expressions.
%However, for the purpose of the present paper, we do not need their explicit expressions.
We have only one term that scales as $\sim \nabla^4\varepsilon^2$,
$c_{10}(\nabla^2\pi)^2$,
and all the other terms scale as
$\sim \nabla^{2(n-1)}\varepsilon^n$ with $n=2,3,$ and 4.
The latter terms are essential in the Vainshtein regime in Horndeski and DHOST theories,
while, as we will see, the former term plays a crucial role in U-DHOST theories with $\alpha_L\neq 0$.

We assume that $M$ is of order of the Planck mass $\mpl$.
We also assume that $\dot\phi_0=\mathcal{O}(MH_0)$, where
$H_0$ is the Hubble parameter today. This is a natural assumption given that
$\phi$ is supposed to be a dark energy field.
It then follows from the explicit expressions that the orders of magnitude for the coefficients are given as in Table~\ref{table:order-list}.

%Although we have not presented the explicit expressions for 
%the coefficients $c_1$, $c_2$, $c_3$, and $b_1$, we may naturally assume that 
%$c_1\sim c_2 =\mathcal{O}(M^{-1})$, $c_3=\mathcal{O}(M^{-2})$, and $b_1=\mathcal{O}(M^{-3}H_0^{-2})$, as in Horndeski/DHOST dark energy models.
%All the EFT parameters are assumed to be (less than) of order $\mathcal{O}(1)$.

\section{Linear regime in theories with $a_1+a_2\neq 0$}\label{sec:LR}

Let us first consider static gravitational fields
in the linear regime in theories with $a_1+a_2\neq 0$ (and hence $\alpha_L\neq 0$)
by dropping all the nonlinear interaction terms in Eq.~\eqref{effLag}.
The linear equations of motion read 
\begin{align}
        c_1\nabla^2\pi +c_4\nabla^2\Psi +2c_6\nabla^2\Phi &= \frac{2}{M^2}\rho,
        \\ 
        c_2\nabla^2\pi+c_4\nabla^2\Phi+2c_5\nabla^2\Psi &=0,
        \\ 
        c_1\nabla^2\Phi+c_2\nabla^2\Psi+2c_3\nabla^2\pi + 2c_{10}\nabla^4\pi&=0,\label{eq-pi-lin}
\end{align}
where we set $a=1$.
After simple algebra we have
\begin{align}
    \left[-c_1c_2c_4+c_1^2c_5+c_2^2c_6+c_3(c_4^2-4c_5c_6)+
    (c_4^2-4c_5c_6)c_{10}\nabla^2\right]\nabla^2 
        \left(
        \begin{array}{c}
        \Phi \\
        \Psi \\
        \pi
        \end{array}
        \right)=
        \left(
        \begin{array}{c}
        c_2^2-4c_5(c_3+c_{10}\nabla^2) \\
        -c_1c_2+2c_4(c_3+c_{10}\nabla^2) \\
        2c_1c_5-c_2c_4
        \end{array}
        \right)\frac{\rho}{M^2}.\label{pppeq}
\end{align}
It can be seen from the orders of magnitude of the coefficients that
\begin{align}
    \underbrace{-c_1c_2c_4+c_1^2c_5+c_2^2c_6+c_3(c_4^2-4c_5c_6)}_{\displaystyle{=\mathcal{O}(M^{-2})}}
    +
    \underbrace{(c_4^2-4c_5c_6)c_{10}\nabla^2}_{\displaystyle{=\alpha_L\times\mathcal{O}(M^{-2}H_0^{-2}\nabla^2)}}
    \simeq (c_4^2-4c_5c_6)c_{10}\nabla^2
\end{align}
on subhorizon scales satisfying $\alpha_L\nabla^2\gg H_0^2$.
Similarly, we have
\begin{align}
    c_2^2-4c_5(c_3+c_{10}\nabla^2)&\simeq -4c_5c_{10}\nabla^2,
    \\ 
    -c_1c_2+2c_4(c_3+c_{10}\nabla^2)&\simeq 2c_4c_{10}\nabla^2.
\end{align}
Using the expressions for the coefficients in terms of the EFT parameters,
we thus obtain
\begin{align}
        \nabla^2\Phi &\simeq 4\pi G_{N}^{(\alpha_L\neq 0)}\rho ,
        \\
        \Psi &\simeq \left(\frac{1+\alpha_H}{1+\alpha_T}\right) \Phi,
        \\ 
        \alpha_L\nabla^4\pi&\simeq \left[
                \frac{c_1(1+\alpha_T)+c_2(1+\alpha_H)}{2(1+\alpha_H)^2-(1+\alpha_T)\beta_3}
        \right]\frac{3X\rho}{2M^2},
\end{align}
where 
\begin{align}
        8\pi G_N^{(\alpha_L\neq 0)}:=
                \left[\frac{2(1+\alpha_T)}{2(1+\alpha_H)^2-(1+\alpha_T)\beta_3}\right]\frac{1}{M^2}.
\end{align}
That $\pi$ obeys the fourth-order differential equation with respect to the spatial coordinates signals the presence of the instantaneous mode.
Noting that $c_1\sim c_2=\mathcal{O}(M^{-1})$, we have 
\begin{align}
        \alpha_L\frac{\nabla^2\pi}{M}\sim \frac{X}{M^2\nabla^2}\cdot\nabla^2\Phi
        \sim \frac{H_0^2}{\nabla^2}\cdot\nabla^2\Phi \ll\nabla^2\Phi,
\end{align}
and we can thus check that
the fluctuation of the scalar field is suppressed compared to the gravitational potential.
As far as the metric perturbations $\Phi$ and $\Psi$ on small scales are concerned,
we do not need the explicit expressions for $c_1$, $c_2$, and $c_3$ in order to give a prediction.

The above expressions themselves are not particularly new, as
they were already obtained in Ref.~\cite{Langlois:2017mxy}.
However, Ref.~\cite{Langlois:2017mxy} focuses on the case of totally degenerate theories and it is argued that if the class IIa degeneracy conditions are satisfied
then the effective gravitational coupling $G_N^{(\alpha_L\neq 0)}$ diverges,
as seen from Eq.~\eqref{deg-IIa-para}.
In contrast, in U-DHOST theories,
$G_N^{(\alpha_L\neq 0)}$ is finite and the standard Newtonian behavior of
gravity is reproduced thanks to
the breaking of the degeneracy conditions.
Note, however, that we still have $\Psi\neq\Phi$ in general, and
the parametrized post-Newtonian parameter $\gamma=\Psi/\Phi$ is
given by $\gamma=(1+\alpha_H)/(1+\alpha_T)$, which must satisfy
$|\gamma-1|\lesssim 10^{-5}$ according to the experimental constraints
in the Solar System~\cite{Will:2018bme}.
Therefore, only the subset of theories satisfying $|\alpha_H-\alpha_T|\lesssim 10^{-5}$ is
phenomenologically acceptable.

As opposed to the Vainshtein regime in Horndeski and DHOST theories, 
the linear study is sufficient in the present analysis of U-DHOST theories with $\alpha_L\neq 0$.
For example, one could have nonlinear terms of order $\mathcal{O}(b_1(\nabla^2\pi)^2)$ and $\mathcal{O}(d_1(\nabla^2\pi)^3)$
in the $\pi$-equation of motion~\eqref{eq-pi-lin}. However, it is easy to see that
the former and latter terms are suppressed respectively by
$(H_0^2/\nabla^2)\Phi$ and $(H_0^2/\nabla^2)\Phi^2$ compared to the linear terms.
Thus, the linear result here is not intervened by the nonlinear derivative interaction terms
that are relevant to the Vainshtein mechanism in Horndeski and DHOST theories.
In other words, the Vainshtein mechanism does not work to make $\Psi=\Phi$,
and as a result $\alpha_T$ and $\alpha_H$ are indeed constrained by the experiments
in the Solar System as stated above.

Even if one sticks to theories with $c_{\textrm{GW}}=1$ and no deviations from
the standard GR result in the Solar System,\footnote{Note, however, that
here we have only evaluated the PPN parameter $\gamma$. It would be interesting to
calculate all the PPN parameters in U-DHOST theories~\cite{Saito:2024xdx}.} a large theory space is still allowed.
In terms of the functions in the Lagrangian, the condition $\alpha_H=\alpha_T=0$ implies
\begin{align}
    f=f(\phi),\quad a_1=\frac{f(\phi)}{2X},
\end{align}
but $a_2$, $a_3$, and $a_4$ are free as long as $a_2\neq -a_1$.

So far we have considered static gravitational fields, but
the result can be extended straightforwardly to allow for time dependence.
Reviving the scale factor, it is easy to see that even in a time-dependent setup one has 
\begin{align}
        \frac{\nabla^2\Phi}{a^2}=
        \left(\frac{1+\alpha_T}{1+\alpha_H}\right)
        \frac{\nabla^2\Psi}{a^2}=4\pi G_N^{(\alpha_L\neq 0)}\rho(t,\Vec{x}).
\end{align}
This result is valid in the limit $\nabla^2\gg a^2H_0^2$ and
can be used for studying the evolution of cosmological density perturbations at deep subhorizon scales.
Note, however, that there must be deviations from GR on large scales even in
theories with $\alpha_T=\alpha_H$
(in addition to a possible time dependence of $G_N^{(\alpha_L\neq 0)}$),
as seen from Eq.~\eqref{pppeq}.
Cosmological tests of such theories would therefore be intriguing.

\section{Vainshtein regime in theories with $a_1+a_2=0$}\label{sec:VR}

\subsection{Modified gravitational forces}

Let us now turn to the case where one of the class Ia degeneracy conditions
is satisfied, $a_1+a_2=0$,
but another condition~\eqref{deg-Ia-2} is not satisfied.
In this case, we have $\alpha_L=\alpha_{V1}-\alpha_{V2}=0$, and hence
the higher-derivative term with the coefficient $c_{10}$ which played an essential role
in the previous section vanishes.\footnote{In principle, one can consider a background satisfying
$\alpha_L=0$ and $\alpha_{V1}\neq\alpha_{V2}$ by carefully tuning the model so that
$a_1(\dot\phi_0^2/2)=a_2(\dot\phi_0^2/2)$ and $a_{1X}(\dot\phi_0^2/2)\neq a_{2X}(\dot\phi_0^2/2)$.
We do not study this case.}
Upon substituting $\alpha_L=\alpha_{V1}-\alpha_{V2}=0$,
the apparent form of the Lagrangian~\eqref{effLag} is identical to that obtained in quadratic
DHOST theories~\cite{Dima:2017pwp,Langlois:2017dyl}, with
all the terms being of order $\mathcal{O}(\nabla^{2(n-1)}\varepsilon^n)$,
which implies that the nonlinear derivative interaction terms are no longer negligible.
The difference from the previous analysis of totally degenerate theories
is that in U-DHOST theories $\beta_3$ is an independent parameter
that is free of the degeneracy conditions~\eqref{deg-Ia-para}.
As we will see below, this drastically modifies the gravitational forces around a source
in contrast to the case of the Horndeski and DHOST theories.

We focus on the nonlinear Vainshtein regime in which we have
\begin{align}
        \frac{\nabla^2\Phi}{a^2}\sim
        \frac{\nabla^2\Psi}{a^2}\sim \frac{(\nabla^2\pi)^2}{a^4X}\sim \frac{\rho}{M^2}
        \gg \frac{\nabla^2\pi}{Ma^2}.\label{def:Vainshtein-regime}
\end{align}
One can estimate that this approximation is valid on scales smaller than
the Vainshtein radius $(r_SH_0^{-2})^{1/3}$,
where $r_S$ is the Schwarzschild radius of the gravitational source.
We consider a static and spherically symmetric setup
in the Vainshtein regime. Setting again $a=1$ and using $r=|\Vec{x}|$,
the equations of motion take the form 
\begin{align}
        \frac{\D\mathcal{J}_\Phi}{\D r}=\frac{\D\mathcal{J}_\Psi}{\D r}=\frac{\D\mathcal{J}_\pi}{\D r}=0,
\end{align}
where
\begin{align}
        \mathcal{J}_\Phi&:=
        -\left(\alpha_H-\alpha_{V}+4\beta_1\right)r(\Pi')^2
        -\left(2\beta_1+\beta_3\right)r^2\Pi'\Pi'' 
        -\beta_3r^2\Phi'+2(1+\alpha_H)r^2\Psi'-\frac{\mathcal{M}(r)}{4\pi M^2},
        \\ 
        \mathcal{J}_\Psi&:= \alpha_Tr(\Pi')^2+2\alpha_H r^2\Pi'\Pi'' 
        +2(1+\alpha_H)r^2\Phi'-2(1+\alpha_T)r^2\Psi',
        \\ 
        \mathcal{J}_\pi&:=2\left(-\alpha_H+\alpha_{V}-2\beta_1+\beta_3\right)
        r\Pi'\Phi'+\left(2\beta_1+\beta_3\right)r^2\Pi'\Phi''
        +2\left(\alpha_T-2\alpha_H \right)r\Pi'\Psi'
        \notag \\ & \quad 
        -2\alpha_H r^2\Pi'\Psi''
        +\left(\alpha_H-\alpha_T-\alpha_{V}+4\beta_1\right)(\Pi')^3
        +(4\beta_1+\beta_3) \left[r\Pi'\Pi'''+r(\Pi'')^2+2\Pi'\Pi''\right]r\Pi',
\end{align}
and we introduced $\alpha_V:=\alpha_{V1}=\alpha_{V2}$, $\Pi:=\pi/\sqrt{2X}$, and 
the enclosed mass inside a sphere of radius $r$:
\begin{align}
        \mathcal{M}(r):=4\pi\int^r_0\rho(x)x^2\D x.
\end{align}
Here a prime stands for differentiation with respect to $r$.
Regularity at $r=0$ fixes the integration constants, leading to
\begin{align}
        \mathcal{J}_\Phi=\mathcal{J}_\Psi=\mathcal{J}_\pi=0.
\end{align}

To make the difference from DHOST theories explicit, we introduce the new parameter $\xi\,(\neq 0)$ defined by
\begin{align}
        \beta_3=-2\beta_1[2(1+\alpha_H)+(1+\alpha_T)\beta_1]+\xi,
\end{align}
which characterizes the breaking of one of the degeneracy conditions~\eqref{deg-Ia-para}.
Hereafter we consider the theories with $c_{\textrm{GW}}=1$ by setting
\begin{align}
        \alpha_T=0,\quad \alpha_V=-\alpha_H.
\end{align}
This is the phenomenologically most interesting case.
Moreover, with this restriction, we have less complicated equations while we do not lose the
essential physics of U-DHOST theories.
From $\mathcal{J}_\Phi=0$ and $\mathcal{J}_\Psi=0$,
we obtain 
\begin{align}
        \left[2(1+\alpha_H+\beta_1)^2-\xi\right]r^2\Phi'&=
        \frac{\mathcal{M}}{4\pi M^2}+2(\alpha_H+2\beta_1)r\mathcal{X} 
        -\frac{1}{2}\left[2(\alpha_H+\beta_1)(1+\alpha_H+\beta_1)-\xi\right]r^2\mathcal{X}',
        \label{soln-Phi}
        \\ 
        \left[2(1+\alpha_H+\beta_1)^2-\xi\right]r^2\Psi'&=(1+\alpha_H)
        \frac{\mathcal{M}}{4\pi M^2}+2(1+\alpha_H)(\alpha_H+2\beta_1)r\mathcal{X} 
        -\frac{1}{2}\left[2\beta_1(1+\alpha_H+\beta_1)-\xi\right]r^2\mathcal{X}',
        \label{soln-Psi}
\end{align}
where $\mathcal{X}:=(\Pi')^2\,(\ge 0)$.
Equivalently, one can write
\begin{align}
    \mathcal{X}&=-\frac{\mathcal{M}/8\pi M^2}{(\alpha_H+2\beta_1)r}
    \notag \\ & \quad 
    +\frac{r}{\alpha_H(\alpha_H+2\beta_1)}
    \left\{
        \left[(\alpha_H+\beta_1)(1+\alpha_H+\beta_1)-\frac{\xi}{2}\right]\Psi' 
        -\left[\beta_1(1+\alpha_H+\beta_1)-\frac{\xi}{2}\right]\Phi'
    \right\},\label{solnX}
    \\ 
    \mathcal{X}'&=\frac{2}{\alpha_H}\left[\Psi'-(1+\alpha_H)\Phi'\right].
    \label{solndX}
\end{align}
These equations can be used to eliminate $\Phi'$, $\Psi'$, $\Phi''$, and $\Psi''$
from $\mathcal{J}_\pi$, yielding 
\begin{align}
        \mathcal{J}_\pi\propto \Pi' 
        \left[
                \xi r^2 \mathcal{X}''+2\xi r\mathcal{X}'+4(\alpha_H+2\beta_1)(1-\alpha_H-3\beta_1)
                \mathcal{X}-\mathcal{S}
        \right]=0,\label{Jpi-fin}
\end{align}
where
\begin{align}
        \mathcal{S}:=4(\alpha_H+2\beta_1)\frac{\mathcal{M}}{4\pi M^2r} 
        +\left[
                2(\alpha_H+\beta_1)(1+\alpha_H+\beta_1)-\xi 
        \right]\frac{\mathcal{M}'}{4\pi M^2}.
\end{align}
We discard the solution $\Pi'=0$, because
this branch does not satisfy~\eqref{def:Vainshtein-regime} and
hence is not the solution we are looking for in the nonlinear regime.
We therefore consider the branch in which the expression inside
the square brackets in Eq.~\eqref{Jpi-fin} vanishes:
\begin{align}
        \xi r^2 \mathcal{X}''+2\xi r\mathcal{X}'+4(\alpha_H+2\beta_1)(1-\alpha_H-3\beta_1)
                \mathcal{X}=\mathcal{S}.\label{eq-for-calX}
\end{align}
Now the particular nature of U-DHOST theories is manifest.
If $\xi=0$, this reduces to an algebraic equation giving
the Vainshtein solution obtained in the previous
studies of DHOST theories~\cite{Kobayashi:2014ida,Dima:2017pwp,Langlois:2017dyl}.
If $\xi\neq 0$, Eq.~\eqref{eq-for-calX} is a differential equation signaling
the presence of an instantaneous mode.

One can obtain the general solution to Eq.~\eqref{eq-for-calX}.
Substituting then the solution to Eqs.~\eqref{soln-Phi} and~\eqref{soln-Psi},
one can express $\Phi'$ and $\Psi'$ in terms of $\mathcal{M}$.
However, to contrast our result with the previous one in DHOST theories~\cite{Dima:2017pwp,Langlois:2017dyl},
it is more convenient to proceed in a different way and derive a single third-order differential equation for the potential $\Phi$.
This can be done as follows.
Using Eq.~\eqref{soln-Phi}, its derivative, and Eq.~\eqref{eq-for-calX}, one can express $\mathcal{X}$
and $\mathcal{X}'$ in terms of $\Phi'$ and $\Phi''$:
$\mathcal{X}=(\dots)\Phi'+(\dots)\Phi''$,
$\mathcal{X}'=(\dots)\Phi'+(\dots)\Phi''$.
Equating the derivative of the former to the latter, we obtain the following third-order equation for $\Phi$:
\begin{align}
    \Xi_*\sigma''+\frac{\sigma}{r^2}
    =
    \Xi_1G_N^{(\alpha_L=0)}\mathcal{M}'',\label{eq:3Phi}
\end{align}
where
\begin{align}
    \sigma:=r^2\Phi'-G_N^{(\alpha_L=0)}\mathcal{M},
\end{align}
and 
\begin{align}
    8\pi G_N^{(\alpha_L=0)}&:=\frac{1}{M^2(1-\alpha_H-3\beta_1)},
    \\ 
    \Xi_*&:=\frac{\xi}{4(1-\alpha_H-3\beta_1)(\alpha_H+2\beta_1)},
    \\
    \Xi_1&:=-\frac{(\alpha_H+\beta_1)^2}{2(\alpha_H+2\beta_1)}
    -\Xi_*(\alpha_H+3\beta_1).
\end{align}
We then eliminate $\mathcal{X}$ and $\mathcal{X}'$ from Eq.~\eqref{soln-Psi} to get
\begin{align}
    \Psi'&=
    \frac{G_N^{(\alpha_L=0)}\mathcal{M}}{r^2}+
    \alpha_H\frac{G_N^{(\alpha_L=0)}\mathcal{M}'}{r}
    -\frac{\beta_1(\alpha_H+\beta_1)}{2(\alpha_H+2\beta_1)}G_N^{(\alpha_L=0)}\mathcal{M}''
    \notag \\ & \quad 
    -\frac{\Xi_*}{\Xi_1}
    \left[(\alpha_H+3\beta_1)\frac{\sigma}{r^2}+
    \alpha_H\frac{\sigma'}{r}
    -\frac{\beta_1(\alpha_H+\beta_1)}{2(\alpha_H+2\beta_1)}\sigma''
    \right].\label{eq:3Psi}
\end{align}
In the case of $\xi=0$, Eqs.~\eqref{eq:3Phi} and~\eqref{eq:3Psi} reproduce the result obtained in
Refs.~\cite{Dima:2017pwp,Langlois:2017dyl}.

The general solution to Eq.~\eqref{eq:3Phi} is obtained as
\begin{align}
    \sigma&=\frac{\Xi_1G_N^{(\alpha_L=0)}}{\Xi_*(n_+-n_-)}\left[
        r^{n_+}\int^r_{R_0}x^{n_-}\mathcal{M}''(x)\D x 
        -r^{n_-}\int^r_{R_0}x^{n_+}\mathcal{M}''(x)\D x 
    \right]
    \notag \\ &\quad 
    +\mathcal{C}_1G_N^{(\alpha_L=0)}\mathcal{M}_0\left(\frac{r}{R_0}\right)^{n_+}
    +\mathcal{C}_2G_N^{(\alpha_L=0)}\mathcal{M}_0\left(\frac{r}{R_0}\right)^{n_-},\label{soln:sigma1}
\end{align}
where
\begin{align}
    n_\pm:=\frac{1}{2}\left(1\pm\sqrt{1-4/\Xi_*}\right),
\end{align}
and $\mathcal{C}_1$ and $\mathcal{C}_2$ are integration constants.
In the first line, the lower bounds of the integrals are arbitrary,
but for convenience, we set them to be the radius of the spherical object $R_0$.
Since $\mathcal{M}''=0$ for $r>R_0$, the first line vanishes in the region exterior to the object.
In the second line,
we introduced $R_0$ and the total mass $\mathcal{M}_0:=\mathcal{M}(R_0)$
so that the integration constants $\mathcal{C}_{1}$ and $\mathcal{C}_2$ are dimensionless.
Note that $1-4/\Xi_*$ can be negative,
and in that case $\sqrt{1-4/\Xi_*}$ in the equations here and hereafter
should be replaced by $i\sqrt{|1-4/\Xi_*|}$.

In the following analysis,
we assume that $\alpha_H$, $\beta_1$, and $\xi$ are small numbers of
$\mathcal{O}(\epsilon)$. It then follows that $\Xi_*=\mathcal{O}(1)$ and
$\Xi_1=\mathcal{O}(\epsilon)$. This assumption is used to estimate the
corrections to the standard result in general relativity.
If $\sigma$ is suppressed by a factor of order $\mathcal{O}(\epsilon)$,
so is $\Psi'-G_N^{(\alpha_L=0)}\mathcal{M}/r^2$, and the deviation from general relativity is expected to be small.

In the region exterior to the spherical object, we have $\mathcal{M}''=0$, and hence
\begin{align}
    %\sigma = \mathcal{C}_1G_N^{(\alpha_L=0)}\mathcal{M}_0\left(\frac{r}{R_0}\right)^{n_+}
    %+\mathcal{C}_2G_N^{(\alpha_L=0)}\mathcal{M}_0\left(\frac{r}{R_0}\right)^{n_-}.
    \Phi' = \frac{G_N^{(\alpha_L=0)}\mathcal{M}_0}{r^2}
    \left[
        1+\Delta_\Phi(r)
        %+\mathcal{C}_1\left(\frac{r}{R_0}\right)^{n_+}
        %+\mathcal{C}_2\left(\frac{r}{R_0}\right)^{n_-}
    \right],\label{extsolPhi}
        \\ 
    \Psi' = \frac{G_N^{(\alpha_L=0)}\mathcal{M}_0}{r^2}
    \left[
        1+\Delta_\Psi(r)
        %+\widetilde{\mathcal{C}}_1\left(\frac{r}{R_0}\right)^{n_+}
        %+\widetilde{\mathcal{C}}_2\left(\frac{r}{R_0}\right)^{n_-}
    \right],\label{extsolPsi}
\end{align}
where 
\begin{align}
    \Delta_\Phi(r)&:=\mathcal{C}_1\left(\frac{r}{R_0}\right)^{n_+}
        +\mathcal{C}_2\left(\frac{r}{R_0}\right)^{n_-},
        \\
    \Delta_\Psi(r)&:=\widetilde{\mathcal{C}}_1\left(\frac{r}{R_0}\right)^{n_+}
        +\widetilde{\mathcal{C}}_2\left(\frac{r}{R_0}\right)^{n_-},
\end{align}
with 
\begin{align}
    \widetilde{\mathcal{C}}_1&=-\frac{\Xi_*}{\Xi_1}\left[
        (1+n_+)\alpha_H+3\beta_1+
        \frac{1}{\Xi_*}\frac{\beta_1(\alpha_H+\beta_1)}{2(\alpha_H+2\beta_1)}
    \right]\mathcal{C}_1,
    \\ 
    \widetilde{\mathcal{C}}_2&=-\frac{\Xi_*}{\Xi_1}\left[
        (1+n_-)\alpha_H+3\beta_1+
        \frac{1}{\Xi_*}\frac{\beta_1(\alpha_H+\beta_1)}{2(\alpha_H+2\beta_1)}
    \right]\mathcal{C}_2.
\end{align}
Substituting Eqs.~\eqref{extsolPhi} and~\eqref{extsolPsi}
to Eq.~\eqref{solnX}, we obtain
\begin{align}
    \mathcal{X}=\frac{2G_N^{(\alpha_L=0)}\mathcal{M}_0}{r}
    \left[
        1-\frac{2\beta_1(1+\alpha_H+\beta_1)-\xi}{4\alpha_H(\alpha_H+2\beta_1)}\Delta_\Phi 
        +\frac{2(\alpha_H+\beta_1)(1+\alpha_H+\beta_1)-\xi}{4\alpha_H(\alpha_H+2\beta_1)}
        \Delta_\Psi 
    \right].\label{solnX-explicit}
\end{align}

To see the difference from the result in the Horndeski and DHOST theories, i.e.,
the deviation from the standard (post-)Newtonian behavior, it is necessary to determine
the integration constants $\mathcal{C}_1$ and $\mathcal{C}_2$.
For this purpose, let us turn to discuss the boundary conditions.

In the vicinity of the center, we have $\mathcal{M}\simeq (4\pi/3)\rho(0)r^3$, and hence
it follows from Eq.~(\ref{soln:sigma1}) that
\begin{align}
    \sigma \simeq \left(\frac{\Xi_1}{1+6\Xi_*}\right)\cdot 8\pi G_N^{(\alpha_L=0)}\rho(0)r^3
    +\mathcal{D}_1G_N^{(\alpha_L=0)}\mathcal{M}_0\left(\frac{r}{R_0}\right)^{n_+}
    +\mathcal{D}_2G_N^{(\alpha_L=0)}\mathcal{M}_0\left(\frac{r}{R_0}\right)^{n_-},
\end{align}
with 
\begin{align}
    \mathcal{D}_{1}&:=\mathcal{C}_{1}-
    \frac{\Xi_1}{\Xi_*(n_+-n_-)}\frac{R_0^{n_+}}{\mathcal{M}_0}
    \int^{R_0}_0x^{n_- }\mathcal{M}''(x)\D x,
    \\
    \mathcal{D}_{2}&:=\mathcal{C}_{2}+ 
    \frac{\Xi_1}{\Xi_*(n_+-n_-)}\frac{R_0^{n_-}}{\mathcal{M}_0}
    \int^{R_0}_0x^{n_+ }\mathcal{M}''(x)\D x.
\end{align}
Let us consider the following three cases:
(i) $0<\Xi_*<4$ ($n_\pm=(1/2)\left(1\pm i\sqrt{|1-4/\Xi_*|}\right)$);
(ii) $\Xi_*<-1/6$ or $\Xi_*>4$ ($1/2<n_+<3$, $-2<n_-<1/2$);
and (iii) $-1/6<\Xi_*<0$ ($n_+>3$, $n_-<-2$).
At the center, we impose that the curvature tensors do not diverge.
Since the curvature tensors $\supset \Phi'',\Psi''\sim \sigma/r^3$, this amounts to requiring that
$\mathcal{D}_1r^{n_+-3}$ and $\mathcal{D}_2r^{n_--3}$ do not diverge as $r\to 0$.
Therefore, in all the above cases one can fix the integration constant $\mathcal{C}_2$ as
\begin{align}
    \mathcal{D}_2=0\quad\Rightarrow\quad
    \mathcal{C}_2=-\frac{\Xi_1}{\Xi_*(n_+-n_-)}\frac{R_0^{n_-}}{\mathcal{M}_0}
    \int^{R_0}_0x^{n_+ }\mathcal{M}''(x)\D x.
\end{align}
Furthermore, in cases (i) and (ii), one can fix $\mathcal{C}_1$ as
\begin{align}
    \mathcal{D}_1=0\quad\Rightarrow\quad 
    \mathcal{C}_{1}=
    \frac{\Xi_1}{\Xi_*(n_+-n_-)}\frac{R_0^{n_+}}{\mathcal{M}_0}
    \int^{R_0}_0x^{n_- }\mathcal{M}''(x)\D x.
\end{align}
The precise values of the integration constants depend on the detailed information of the density profile.
However, one can roughly estimate their size as $\mathcal{C}_{1,2} =\mathcal{O}(\epsilon)$,
and hence the correction terms $\Delta_\Phi$ and $\Delta_\Psi$ are suppressed by $\epsilon$ if the EFT parameters are as small as $\mathcal{O}(\epsilon)$.

Having determined the integration constants $\mathcal{C}_{1,2}$ in cases (i) and (ii),
we now see that
in case (i) the gravitational potentials have the correction terms showing an oscillatory behavior.
The correction terms grow as
\begin{align}
    \Delta_\Phi,\Delta_\Psi\sim
    \mathcal{O}(\epsilon)\times \sqrt{\frac{r}{R_0}}\cos\left(\frac{1}{2}\sqrt{|1-4/\Xi_*|}\ln r\right),
    \quad \mathcal{O}(\epsilon)\times \sqrt{\frac{r}{R_0}}
    \sin\left(\frac{1}{2}\sqrt{|1-4/\Xi_*|}\ln r\right),
\end{align}
and would eventually dominate over the Newtonian term.
What is more problematic is that these oscillatory corrections
appear as
$\mathcal{O}(\epsilon^{-1})\times \Delta_\Phi,\mathcal{O}(\epsilon^{-1})\times\Delta_\Psi$
in Eq.~\eqref{solnX-explicit} and thus the second and third terms easily make $\mathcal{X}$ negative. Recalling that $\mathcal{X}=(\Pi')^2$ must be non-negative, case (i) is not acceptable
(except for the case where the parameters are fine-tuned so that the coefficients
of the second and third terms in Eq.~\eqref{solnX-explicit} vanish).

%Therefore,
%$\mathcal{C}_{1,2}$ must be sufficiently small so that the corrections remain small, e.g., in the Solar System.
%(Setting $R_0=R_\odot$, the factor $\sqrt{r/R_\odot}$ evaluated at the orbit of Mercury (Neptune) is given by $\sim 10 \,(80)$, and hence the enhancement is not actually that significant.)

In case (ii), the correction terms in Eqs.~\eqref{extsolPhi} and~\eqref{extsolPsi}
grow as $\sim r^{n_+}$ with $1/2<n_+<3$ and dominate immediately over the standard Newtonian forces.
Therefore, this case is also dangerous.

In contrast, in case (iii), $\Phi''$ and $\Psi''$ do not diverge at the center even if $\mathcal{D}_1\neq 0$.
Instead, one can set $\mathcal{C}_1=0$ so that the correction terms fall off as $\sim r^{n_-}$ with $n_-<-2$. Therefore, there are small corrections to the gravitational forces and they remain small:
\begin{align}
    \Delta_\Phi %= -\frac{\Xi_1}{\Xi_*(n_+-n_-)}\frac{1}{\mathcal{M}_0}
    %\int^{R_0}_0x^{n_+}{\mathcal{M}}''\D x\cdot r^{n_-}
    \sim \epsilon\left(\frac{r}{R_0}\right)^{n_-}\quad n_-<-2.
\end{align}
%which is the most interesting case among the three.
For a quasicircular orbit, this modification causes the anomalous perihelion advance per orbit 
\begin{align}
    \delta\varphi = \pi r\left[r^2\left(\Delta_\Phi/r\right)'\right]'=\mathcal{O}( \Delta_\Phi ).
\end{align}
For Mercury we have $\delta\varphi\sim \epsilon \times 10^{2n_-}$,
while the observational error of the advance of Mercury's perihelion is given by 
$\sigma_{\delta\varphi}\sim 10^{-10}$~\cite{Pitjeva:2005},
leading to the bound
$\epsilon \times 10^{2n_-}\lesssim 10^{-10}$ (note that the power $n_-$ depends on the parameters).
Using Moon we obtain a similar bound.

In the interior region, the additional corrections come from
nonvanishing $\mathcal{M}'$ and $\mathcal{M}''$, as seen from Eqs.~\eqref{eq:3Psi} and~\eqref{soln:sigma1}. Note that, for given $\mathcal{M}=\mathcal{M}(r)$, these corrections appear in a different way
from the case of DHOST theories ($\Xi_*=0$), because in the DHOST case
we simply obtain $\sigma=\Xi_1G_N^{(\alpha_L=0)}\mathcal{M}''r^2$
without solving the second-order differential equation for $\sigma$.
In the simple case of constant matter density where $\mathcal{M}(r)=\mathcal{M}_0(r/R_0)^3$ for $r\le R_0$, it follows from Eq.~\eqref{soln:sigma1} that the correction to the Newtonian force in the interior region is given by 
\begin{align}
    \Phi'-\frac{G_N^{(\alpha_L=0)}\mathcal{M}}{r^2}=
    6\Xi_1\frac{G_N^{(\alpha_L=0)}\mathcal{M}}{r^2}\left[
    \frac{1}{1+6\Xi_*}
    -\frac{2}{\sqrt{4-\Xi_*}(5\sqrt{-\Xi_*}+\sqrt{4-\Xi_*})}
    \left(\frac{r}{R_0}\right)^{n_+-3}
    \right].
\end{align}
Here, we presented only the result of case (iii) ($-1/6<\Xi_*<0$, $n_+>3$, $n_-<-2$) since the correction in the exterior region remains small only in this case, as we have seen above.
For completeness, we present the exterior solution explicitly,
\begin{align}
    \Phi'-\frac{G_N^{(\alpha_L=0)}\mathcal{M}_0}{r^2}=
    -\frac{G_N^{(\alpha_L=0)}\mathcal{M}_0}{r^2}\cdot 
    \frac{12\Xi_1}{\sqrt{4-\Xi_*}(5\sqrt{-\Xi_*}-\sqrt{4-\Xi_*})}
    \left(\frac{r}{R_0}\right)^{n_-}.
\end{align}
Note that $\Phi'$ is continuous across $r=R_0$.

\section{Conclusions}\label{sec:concl}

Higher-order scalar-tensor theories in general have unstable Ostrogradsky modes due to the higher-derivative nature of the field equations. To evade this instability, the degeneracy conditions are imposed among functions in the Lagrangian so that the kinetic matrix is degenerate in any gauge. The dangerous Ostrogradsky mode is thus removed in such (totally) degenerate higher-order scalar-tensor (DHOST) theories~\cite{Langlois:2015cwa,Crisostomi:2016czh,BenAchour:2016fzp}.
Recently, it was noticed that one can have healthy higher-order scalar-tensor theories even if the degeneracy conditions are satisfied only in the unitary gauge and are thus broken partially.
A seemingly dangerous mode then appears, but it is an instantaneous mode obeying an elliptic equation on spacelike hypersurfaces, and hence the system is free from instabilities under appropriate boundary conditions. Such theories are called U-DHOST theories, whose action is given by Eq.~\eqref{action:udhost} with Eqs.~\eqref{lag:sup1} and~\eqref{lag:sup2}.

In this paper, we have studied weak gravitational fields produced by a nonrelativistic source in U-DHOST theories.
We have paid particular attention to the role of nonlinear derivative interactions that are essential for the Vainshtein screening mechanism in the Horndeski theory~\cite{Kimura:2011dc,Narikawa:2013pjr,Koyama:2013paa} and its breaking in theories beyond Horndeski~\cite{Kobayashi:2014ida,Dima:2017pwp,Langlois:2017dyl,Hirano:2019scf,Crisostomi:2019yfo}.

First, we have considered the generic U-DHOST theories satisfying $A_1+A_2=a_1+a_2\neq 0$,
characterized by $\alpha_L\neq 0$ in terms of the so-called EFT parameters.
In sharp contrast to the case of totally degenerate theories,
we have found that nonlinear terms can never be important and the linear analysis is sufficient in the weak-field regime of U-DHOST theories. 
Since the Vainshtein screening mechanism does not come to help at all,
we need to tune to some extent the form of the Lagrangian in order to pass solar-system tests.
We have identified the subset of U-DHOST theories evading the experimental tests, which is characterized by $f=f(\phi), a_1=f/2X\neq -a_2$, with the other functions of $\phi$ and $X$ being free (see Eqs.~\eqref{action:udhost},~\eqref{lag:sup1}, and~\eqref{lag:sup2} for the more explicit form of the action).
Note that, thanks to the detuning of the degeneracy conditions, U-DHOST theories are free from the pathology of the diverging Newton constant and unavoidable gradient instabilities pointed out in totally degenerate theories with $\alpha_L\neq 0$~\cite{Langlois:2017mxy}.

Our results show that there exists a large family of healthy, viable scalar-tensor theories in the U-DHOST class. Then, it would be interesting to explore cosmologies of U-DHOST theories and test them against cosmological observations such as the cosmic microwave background, using, for example, the Boltzmann solver developed for higher-order scalar-tensor theories in Refs.~\cite{Hiramatsu:2020fcd,Hiramatsu:2022ahs,Hiramatsu:2022fgn}. This is left for further study.

We have also investigated the special case where $\alpha_L\propto A_1+A_2=a_1+a_2=0$, but the remaining one of the degeneracy conditions is still violated.
In this case, the nonlinear derivative interactions play a key role in determining the behavior of weak gravitational fields as in the Vainshtein regime of totally degenerate theories.
Technically, the problem essentially reduces to solving an algebraic equation for the first derivative of the fluctuation of the scalar field, $\pi'$, in the case of totally degenerate theories. In contrast, we need to solve a differential equation for $\pi'$ in U-DHOST theories (with $\alpha_L=0$). As a result, we have found that the Vainshtein screening mechanism does not work well both inside and outside a matter object, giving rise to modifications of gravitational forces. The detail of modifications depends on the density profile of matter as well as the theory parameters in a complicated way, and in some cases the modifications are too large at large distances from the source, invalidating the models.

%--- Acknowledgments ---%--- Acknowledgments ---%--- Acknowledgments ---%
\acknowledgments
The work of T.K. was supported by
JSPS KAKENHI Grant No.~JP20K03936 and
MEXT-JSPS Grant-in-Aid for Transformative Research Areas (A) ``Extreme Universe,''
No.~JP21H05182 and No.~JP21H05189.
The work of T.H. was supported by JSPS KAKENHI Grants No.~JP21K03559 and No.~JP23H00110.
%--- Acknowledgments ---%--- Acknowledgments ---%--- Acknowledgments ---%

\appendix

\section{The nonlinear effective action and the EFT parameters}\label{app:eftlag}

Writing the metric as $\D s^2=-N^2\D t^2+\gamma_{ij}(\D x^2+N^i\D t)(\D x^j+N^j\D t)$,
one can express the U-DHOST action~\eqref{action:udhost} in the unitary gauge in terms of the ADM variables as
\begin{align}
    S=\int\D t\D^3x\sqrt{\gamma}N & \biggl\{\widetilde{P}+\widetilde{Q} K+
        \left(f+2XA_1\right)K_i^jK_j^i-\left(f-2XA_2\right)K^2
        %-\frac{2\dot\phi_0}{N}fK
        +fR^{(3)}
        \notag \\ &
        +
        \frac{2\dot\phi_0}{N}\left(-f_X+A_2-XA_3\right)KV 
        +\left[A_1+A_2-2X(A_3+A_4)+4X^2A_5\right]V^2
        \notag \\ & 
        +4X\left(-f_X-A_1+XA_4\right)\mathbf{a}_i\mathbf{a}^i
    \biggr\},
\end{align}
where $X:=\dot\phi_0^2/(2N^2)$, $K_{ij}$ is the extrinsic curvature of spatial hypersurfaces,
$R^{(3)}$ is the three-dimensional Ricci scalar,
\begin{align}
    V:=\frac{1}{N}\partial_t\left(\frac{\dot\phi_0}{N}\right)+
    \frac{\dot\phi_0}{N}\frac{N_i}{N}\partial_iN,
\end{align}
and $\mathbf{a}_i:=\partial_iN/N$.
Here we defined the functions $\widetilde{P}$ and $\widetilde{Q}$ of $t$ and $N$
as 
\begin{align}
    \widetilde{P}:=P+\sqrt{X}\int^{X}\frac{Q_\phi(\phi,y)}{\sqrt{y}}\D y,
    \quad 
    \widetilde{Q}:=-\textrm{sign}(\dot\phi_0)\int^X\sqrt{2y}Q_{y}(\phi,y)\D y
    -\frac{2\dot\phi_0}{N}f_\phi.
\end{align}
In deriving the above expression we used the fact that for any function $F$ of $\phi_0$ and $X$
the following combination is a total divergence:
\begin{align}
    \sqrt{\gamma}N\left[
        \left(F+2XF_X\right)V+\frac{\dot\phi_0}{N}FK+2XF_\phi
    \right].
\end{align}

The above action is then expanded around a flat FLRW background in terms of
the perturbation of the lapse function, $\delta N=N-1$, and
the perturbation of the extrinsic curvature, $\delta K_i^j=K_i^j-H\delta_i^j$.
We obtain the action of the form
\begin{align}
    S=\int\D t\D^3x\sqrt{\gamma}\frac{M^2}{2}&\biggl\{
        (1+\delta N)\left[\delta K_i^j\delta K_j^i-\left(1+\frac{2}{3}\alpha_L\right)\delta K^2\right]
        +(1+\alpha_T)R^{(3)}+4H\alpha_B\delta N\delta K+(1+\alpha_H)\delta NR^{(3)}
        \notag \\ &
        +4\beta_1\delta K\delta V+\beta_2\delta V^2
        +\beta_3\mathbf{a}_i\mathbf{a}^i-\alpha_{V1}\delta N\delta K_i^j\delta K_j^i
        +\alpha_{V2}\delta N\delta K^2+H^2\alpha_K\delta N^2+\dots
    \biggr\},\label{eq:app:eft-action}
\end{align}
where
$\delta V:=\dot N/N-(N^i/N)\partial_iN$ %, \TH{$\mathbf{a}_i := \partial_iN/N$}
and the ellipses represent the terms that are irrelevant for weak fields in the quasistatic regime
(the last term $H^2\alpha_K\delta N^2$ is also irrelevant for that case).
The EFT parameters introduced in the main text are defined as the coefficients in the
above effective action.

Let us comment on the stability of linear perturbations in U-DHOST theories.
The quadratic action for scalar perturbations around a cosmological background
can be derived from Eq.~\eqref{eq:app:eft-action}. It is known that
the scalar perturbations are unstable in fully degenerate theories with
$\alpha_L\neq 0$ in the case where tensor perturbations are
stable, $M^2>0$, $1+\alpha_T>0$~\cite{deRham:2016wji,Langlois:2017mxy}.
This is not always the case if the degeneracy conditions are relaxed.

It is convenient to introduce $\tilde \zeta$ defined as 
\begin{align}
    \tilde \zeta:=\zeta-\beta_1\delta N,
\end{align}
where $\zeta$ is the usual curvature perturbation, $\gamma_{ij}=a^2(t)e^{2\zeta}\delta_{ij}$.
Ignoring the effect of matter fields,
the quadratic action for $\tilde \zeta$ on subhorizon scales
can be written as~\cite{Langlois:2017mxy}
\begin{align}
    S=\int \D t\D^3x\frac{a^3M^2}{2}\left[
    \frac{6(1+\alpha_L)}{\alpha_L}\left(\partial_t\tilde\zeta\right)^2
    -\frac{\tilde B}{a^2}\left(\partial\tilde\zeta\right)^2
    \right],
\end{align}
where $\alpha_L\neq 0$ has been assumed and 
\begin{align}
    \tilde B:=\frac{2(1+\alpha_L)^2[2(1+\alpha_H)^2-(1+\alpha_T)\beta_3]}%
    {4(1+\alpha_L)(1+\alpha_H)\beta_1+2(1+\alpha_T)\beta_1^2+(1+\alpha_L)^2\beta_3}.
\end{align}
Note that $\tilde B$ neither vanishes nor diverges if one relaxes the degeneracy condition on $\beta_3$.
To avoid ghost and gradient instabilities on subhorizon scales we need to require that 
\begin{align}
    \frac{1+\alpha_L}{\alpha_L}>0,\quad \tilde B>0.
\end{align}
The stability conditions for superhorizon perturbations are much more complicated
and can be found in Appendix D of Ref.~\cite{Langlois:2017mxy}.

\section{The coefficients in the effective Lagrangian~\eqref{effLag}}\label{app:cff}

The coefficients in the effective Lagrangian~\eqref{effLag} are given explicitly by 
\begin{align}
        &c_1:= -\frac{2}{\dot\phi_0}\left\{H\left[2\alpha_B-2\alpha_H+2\alpha_L+\beta_3(1+\alpha_M)\right]+\dot\beta_3\right\},
        \notag \\
        &c_2:= \frac{4}{\dot\phi_0}
        \left\{H\left[\alpha_M+\alpha_H(1+\alpha_M)-\alpha_T\right]+\dot\alpha_H\right\},
        \notag \\ 
        &c_3:= -\frac{1}{4X}\biggl\{
        H^2\left[4\alpha_B(1+\alpha_M)-4(\alpha_H+\alpha_M+\alpha_H\alpha_M-\alpha_T)+(1+\alpha_M)^2\beta_3\right]
        \notag \\ &\quad\quad
        +\dot{H}\left(4+4\alpha_B-4\alpha_H+8\alpha_L+\beta_3+\alpha_M\beta_3\right)
        +H\left[4\dot{\alpha}_B-4\dot{\alpha}_H+\beta_3\dot{\alpha}_M+2(1+\alpha_M)\dot{\beta}_3\right] 
        +\ddot{\beta}_3\biggr\}
        \notag \\ &\quad\quad 
        +\left[H(1+\alpha_M)(4\beta_1+\beta_3)+4\dot\beta_1+\dot\beta_3\right] 
        \frac{\ddot\phi_0}{2\dot\phi_0 X} 
        -(4\beta_1+\beta_3)\frac{\ddot\phi_0^2}{4X^2}
        -\frac{\bar\rho+\bar p}{2M^2X},
        \notag \\
        &c_4:=4(1+\alpha_H),
        \quad 
        c_5:=-2(1+\alpha_T),
        \quad 
        c_6:=-\beta_3,
        \notag \\ &
        c_7:=\frac{4}{\dot\phi_0}\left(\alpha_H-\alpha_L\right),
        \quad 
        c_8:=-\frac{2}{\dot\phi_0}(2\beta_1+\beta_3),
        \quad 
        c_9:=\frac{1}{2X}(4\beta_1+\beta_3),
        \quad 
        c_{10}:=-\frac{\alpha_L}{3X},
        \notag \\ &
        b_1:=\frac{1}{6X\dot{\phi}_0}\biggl\{H \left[12\alpha_{B}-3 \alpha_{H} (\alpha_{M}+3)+4 \alpha_{L} (2 \alpha_{M}-3)+3 \alpha_{M} (\alpha_{V1}-8 \beta_{1}-2)+9 \alpha_{T}-3 \alpha_{V1}\right]
        \notag \\ &\quad\quad        
        -3\dot{\alpha}_{H}+8\dot{\alpha}_{L}+3 \left(\dot{\alpha}_{V1}-8 \dot{\beta}_{1}\right)\biggr\}
        +\left[3 (4 \beta_{1}+\beta_{3})-8 \alpha_{L}\right]\frac{\ddot{\phi}_0}{6X^2},
        \notag \\ &
        \widetilde{b}_1:=-\frac{1}{3M^2\dot\phi_0 X}\frac{\D}{\D t}(M^2\alpha_L)
        -\frac{\ddot\phi_0}{4X^2}(\alpha_{V1}-\alpha_{V2}),
        \quad 
        b_2:=
        \frac{1}{2X}\left[-\alpha_H-4\beta_1+\frac{1}{2}(\alpha_{V1}+\alpha_{V2})+\frac{5}{3}\alpha_L\right],
        \notag \\ &
        \widetilde{b}_2:=-\frac{1}{4X}\left(2\alpha_L+\alpha_{V1}-\alpha_{V2}\right),
        \quad 
        b_3:=\frac{1}{2X}\left(\alpha_T-\frac{5}{3}\alpha_L\right),
        \quad 
        \widetilde{b}_3:=-\frac{\alpha_L}{6X},
        \notag \\ &
        b_4:=-\frac{2}{X}\left(\alpha_H-\frac{\alpha_L}{3}\right),
        \quad 
        b_5:=\frac{1}{X}\left(2\beta_1+\beta_3-\frac{4}{3}\alpha_L\right),
        \quad 
        b_6:=-\frac{1}{\dot\phi_0 X}\left(4\beta_1+\beta_3-\frac{4}{3}\alpha_L\right), 
        \notag \\ &
        \widetilde{b}_6:=\frac{1}{2\dot\phi_0 X}\left(\frac{4}{3}\alpha_L+\alpha_{V1}-\alpha_{V2}\right),
        \quad 
        d_1:=-\frac{1}{2X}(b_2+b_3)-\frac{\alpha_L}{6X^2},
        \notag \\ &
        \widetilde{d}_1:=\frac{1}{8X^2}\left(\frac{4}{3}\alpha_L+\alpha_{V1}-\alpha_{V2}\right),
        \quad 
        d_2:=\frac{1}{4X^2}\left(4\beta_1+\beta_3-\frac{4}{3}\alpha_L\right),
\end{align}
where $\alpha_T$, $\alpha_H$, $\beta_1$, $\beta_3$, $\alpha_L$, $\alpha_{V1}$, and $\alpha_{V2}$
are the EFT parameters defined in Eq.~\eqref{defs:eftparas}, 
$\alpha_M$ and $\alpha_B$ are given by 
\begin{align}
    M^2H\alpha_M&= \frac{\D}{\D t}M^2,
    \\ 
    M^2H\alpha_B&= \left(3 f_{X}+2 X f_{XX}+A_1-2A_{2}-2XA_{2X}+3XA_{3}+2X^2A_{3X}-2XA_{4}+4X^2A_5\right)\dot{\phi}_0\ddot{\phi}_0 
 \notag \\ &\quad 
    -2H\left(f_{X}+2 A_{1}+2 XA_{1X}+3 A_{2}+6 X A_{2X}+3 XA_{3}\right)X
 \notag \\ &\quad 
    +\left(f_{\phi}+2Xf_{\phi X}+X Q_{X}\right)\dot{\phi}_0
,   
\end{align}
$H:=\dot a/a$, and we used the homogeneous part of the field equations to simplify the expression for $c_3$ and express it in terms of the background energy density $\bar \rho$ and pressure $\bar p$.

%-----------------------------------------------------------------%
\bibliography{refs}
\bibliographystyle{JHEP}
%-----------------------------------------------------------------%
\end{document}